\documentclass[doublecol]{epl2}
\usepackage{graphicx}

\title{Reentrant Pinning, Dynamic Row Reduction, and Skyrmion Accumulation for Driven Skyrmions in Inhomogeneous Pinning Arrays }
\shorttitle{Reentrant Pinning, Dynamic Row Reduction and Skyrmion Accumulation...}
 
\author{C. Reichhardt\inst{1,2} \and C. J. O. Reichhardt\inst{1,2}}

\institute{
\inst{1} Theoretical Division and Center for Nonlinear Studies,
Los Alamos National Laboratory, Los Alamos, New Mexico 87545, USA\\
\inst{2} Center for Nonlinear Dynamics, Los Alamos National Laboratory, Los Alamos, New Mexico 87545, USA
}

\abstract{
We examine the dynamics of skyrmions in systems with a region of strong pinning coexisting with a pin free region, where the equilibrium state has a uniform skyrmion density.  Under an applied drive, skyrmions accumulate in the pin free region along the edge of the pinned region due to the skyrmion Hall effect. As the drive increases, a series of dynamical structural transitions occur in the flowing skyrmion lattice similar to those observed in the compression dynamics of crystals.  These transitions correspond to reductions in the number of flowing rows of skyrmions due to the collective motion of the skyrmions into the pinned region, and they are accompanied by a series of steps in the velocity force curves.  When the number of pinning sites is sufficiently large, a drive induced pinning effect can occur when the skyrmion Hall effect forces all of the skyrmions to enter the strongly pinned region.  This reentrant pinning effect becomes more pronounced for increasing intrinsic skyrmion Hall angle. }

\begin{document}

\maketitle

\section{Introduction}
Skyrmions are particle like magnetic textures which were initially observed
in neutron scattering \cite{Muhlbauer09} and, shortly afterwards,
directly imaged with Lorentz microscopy \cite{Yu10}. 
Since then,
skyrmions have been identified in an increasing number of materials, including some
which support skyrmions at room temperature
\cite{Nagaosa13,Jiang15,Woo16,Boulle16,Soumyanarayanan17}.
Skyrmions have many similarities to vortex lattices in type-II superconductors,
including the fact that they form a triangular lattice and can be set into motion with an 
applied current \cite{Nagaosa13}. When quenched disorder
is present, there is a finite depinning threshold above which
the skyrmions enter a sliding state
\cite{Schulz12,Iwasaki13,Liang15,Reichhardt15a,Tolley18,Koshibae18}.
One aspect of skyrmions
that is significantly different from superconducting vortices is that the dynamics of 
each skyrmion is strongly influenced by the
Magnus or gyroscopic force
\cite{Nagaosa13,Schulz12,Iwasaki13,Reichhardt15a,Kim19,Brown18,Reichhardt16}. 
In the absence of pinning, the Magnus force causes
the skyrmions to move at an angle with respect to the drive called 
the skyrmion Hall angle, and
when pinning is present, the skyrmions
undergo a spiraling motion in or around the pinning sites \cite{Nagaosa13}.
If there is no disorder,
the skyrmion moves at
the intrinsic skyrmion Hall angle $\theta^{\rm int}_{sk}$ which is independent of the drive;
however, when pinning is present,
the skyrmion Hall angle becomes strongly drive dependent, increasing
with drive from
a value near zero
just above depinning
and eventually saturating at a value
close to the intrinsic skyrmion Hall angle.
The drive dependent skyrmion Hall angle
has been
observed in particle based simulations
\cite{Reichhardt15a,Reichhardt16,Diaz17},
micromagnetic simulations \cite{Legrand17,Kim17,Juge19}
and experiments \cite{Juge19,Jiang17,Litzius17,Woo18}.

Since the skyrmions move at an angle to the drive,
the skyrmion Hall effect can lead to an
accumulation of skyrmions along the edge of a sample.
One of the limitations in
using skyrmions for devices such as race track memory
is that a skyrmion does not simply travel down the race track but
moves toward the track edge,
where it can escape.
When the Magnus force is larger, the distance the skyrmion can travel
down the track is shorter \cite{Iwasaki13a,EverschorSitte18},
so there have been
various efforts to reduce the skyrmion Hall angle \cite{Woo18,EverschorSitte18,Hirata19}.
Another approach for guiding skyrmions is to create
a controlled pinning landscape that
can be used to control the skyrmion motion.
For example,
inhomogeneous pinning or regions with strong pinning coexisting with regions of
low or no pinning
can be created using techniques similar to those
employed for controlling the motion of  vortices in type-II superconductors
with nanostructured pinning arrays
\cite{Harada96,Martin97,Welp05,Trastoy14}.  

Here we use particle  based simulations to study skyrmion dynamics
in a system containing a pinned
region in the form of a stripe coexisting with a
pin-free region,
where the equilibrium state is a uniform skyrmion lattice. 
Under a drive applied parallel to the pinning stripe,
the skyrmions in the pinned region remain immobile 
and the skyrmions in the unpinned region are guided parallel to the drive
by the confining 
effect of the pinned skyrmions,
giving 
a skyrmion Hall angle of zero for an extended range of drives.
This is accompanied by
the accumulation of skyrmions along the
edge of the pinned region, forming a skyrmion density gradient. 
As the drive increases,
the skyrmions in the unpinned region become increasingly compressed,
resulting in 
sudden rearrangements of the skyrmion structure.
Groups of skyrmions enter the
pinned portion of the sample in order to maintain
an integer number of skyrmion rows
flowing in the unpinned regions,
producing stick-slip jumps in the velocity-force curves along with
regions of negative differential conductivity.
The row reduction events in the skyrmion structure resemble those found
for dynamical compression of crystals \cite{McDermott16}
and particle flow in channels \cite{Kokubo04,Besseling99,Besseling05}.   
When the drive is sufficiently large,
the skyrmions can move all the way through the pinned region,
giving a finite skyrmion Hall angle. 
At even higher drives, the skyrmion density becomes uniform again.
If the number of pinning sites is large enough,
it is possible to observe a drive-induced pinning effect in which, as the current
increases,
all the skyrmions become trapped in
the pinned region. 
This reentrant pinning is enhanced for
larger Magnus force since the skyrmions in the unpinned
region are more strongly pressed against the pinned region,
while in the
overdamped or vortex limit,
the lattice compression dynamics and reentrant pinning are lost.  
We map out the dynamics in a series of phase diagrams
and discuses how these effects should be a general feature of skyrmion dynamics
whenever
the pinning is inhomogeneous.

\section{Simulation}

\begin{figure}
  \onefigure[width=\columnwidth]{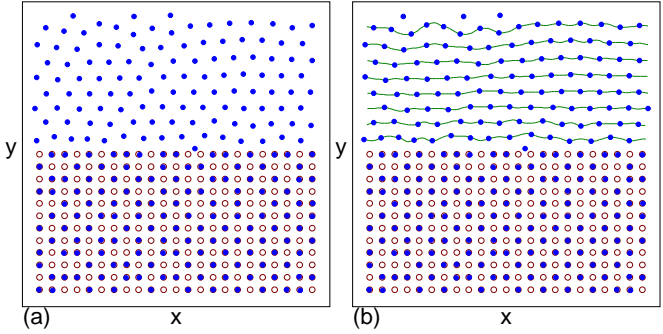}
  \caption{ Skyrmion positions (dots) and pinning sites (open circles) for a system
    with $\alpha_{m}/\alpha_{d} = 1.25$
    containing pinning in only the lower half of the sample.  The drive $F_D$ is
    applied along the $x$ direction.
    The ratio $N_s/N_p$ of skyrmions to pinning sites in the entire sample is 1:1.
    (a) The skyrmion positions after annealing at  $F_{D} = 0$.
    (b) Skyrmion positions and trajectories (lines) at $F_{D} = 0.025$
    showing motion only in the unpinned region.
}
\label{fig:1}
\end{figure}

We consider a two-dimensional system
with periodic boundary conditions in the $x$ and $y$-directions where 
only  half of the sample contains $N_{p}$ pinning sites,
as illustrated in Fig.~\ref{fig:1}(a).
The pinning sites are placed in a square lattice with lattice constant $a$.
The sample contains
$N_{s}$ skyrmions, and the initial state is obtained through simulated annealing.
Due to the repulsive interactions between the skyrmions,
in the absence of a drive the skyrmion density is uniform,
as shown in Fig.~\ref{fig:1}(a).
After annealing, a drive is applied
in the positive $x$-direction.
We use a particle based model for the skyrmion dynamics
in the presence of disorder as employed in previous works
\cite{Reichhardt15a,Brown18,Reichhardt16,Diaz17,Lin13}. 
The skyrmion dynamics is governed by the following equation of motion:      
\begin{equation} 
\alpha_d {\bf v}_{i} + \alpha_m {\hat z} \times {\bf v}_{i} =
{\bf F}^{ss}_{i} + {\bf F}^{D} 
\end{equation}
where ${\bf v}_i$ is the velocity of skyrmion $i$.
The repulsive skyrmion-skyrmion interaction force is  
${\bf F}_{i} = \sum^{N}_{j=1}K_{1}(r_{ij}){\hat {\bf r}_{ij}}$,
where $r_{ij} = |{\bf r}_{i} - {\bf r}_{j}|$, $r_{ij}$ is the distance
between skyrmions $i$ and $j$, and
the modeled Bessel function $K_{1}(r)$ falls
off exponentially for large $r$.
The driving force ${\bf F}^{D}=F^D{\bf {\hat x}}$ is applied uniformly to all
particles in the direction parallel to the stripe of pinning.
We measure the skyrmion velocity both
parallel,
$\langle V_{||}\rangle=N_{s}^{-1}\sum_{i}^{N_s}{\bf v}_i \cdot {\bf \hat x}$,
and perpendicular,
$\langle V_{\perp}\rangle=N_s^{-1}\sum_{i}^{N_s}{\bf v}_i \cdot {\bf \hat y}$,
to the driving force.
We increase the drive
in increments of $\Delta F_D=0.0002$ and
average over 75000 time steps
at each drive to ensure that
the dynamics is in a steady state.  
The damping term $\alpha_d$  aligns the skyrmion velocity
in the direction of the net applied force,
while $\alpha_{m}$ is the coefficient of the Magnus term 
which creates a velocity component perpendicular to the net applied force.
For a finite Magnus term,
in the of absence pinning the skyrmions
move at an angle with respect to the driving force
of $\theta^{\rm int}_{sk} = \arctan(\alpha_{m}/\alpha_{d})$.
In Fig.~\ref{fig:1}, the drive is applied in the positive $x$-direction, and the
resulting skyrmion velocity
is in the positive $x$ and negative $y$ direction. 
The pinning is modeled as parabolic traps of radius $r_{p} = 0.25$
with a maximum strength of $F_{p}$.
The density of the system
is given by $N_s/N_p$,
the ratio of the number of skyrmions in the entire sample to the total number
of pinning sites.
In Fig.~\ref{fig:1}, $N_{s}/N_{p} = 1.0$.
To perform the simulated annealing, we start in
a higher temperature liquid state and gradually cool
the sample to $T = 0.$
Generally this produces
a state with a uniform skyrmion density,
as illustrated in Fig.~\ref{fig:1}(a) for $F_{D} = 0.0$.
Under an applied drive,
the skyrmions in the unpinned region depin first,
as shown in Fig.~\ref{fig:1}(b) at $F_{D} = 0.025$, 
where the skyrmion motion is confined along the $x$ direction.       

\begin{figure}
  \onefigure[width=\columnwidth]{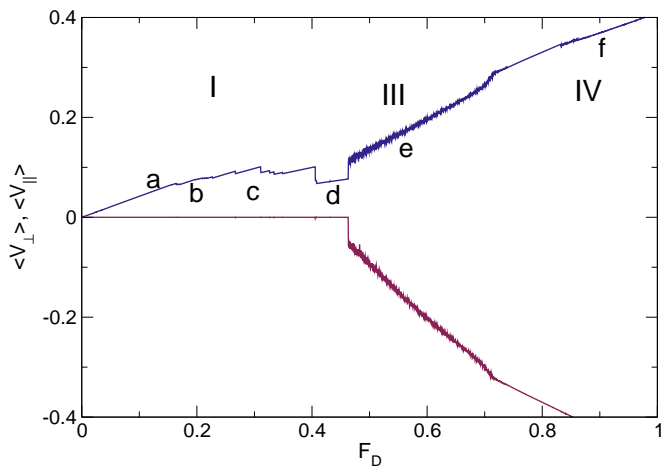}
  \caption{The average skyrmion velocity
    parallel, $\langle V_{||}\rangle$ (blue), and 
    perpendicular, $\langle V_{\perp}\rangle$ (red), to the drive
    vs $F_{D}$ for the system in Fig.~\ref{fig:1} with $F_{p} = 0.75$. 
    Region I is the shear flow where the skyrmions
    in the pin-free region move in the direction of drive,
    as illustrated in Fig.~\ref{fig:3}(a,b,c,d). 
    Region III is the disordered plastic flow where motion
    occurs in both directions, as shown in Fig.~\ref{fig:3}(e),
    and in region IV, all the skyrmions
are moving in a lattice structure as in Fig.~\ref{fig:3}(f).
}
\label{fig:2}
\end{figure}

\section{Shearing Dynamics for Parallel Driving}

In Fig.~\ref{fig:2} we plot $\langle V_{||}\rangle$ and
$\langle V_{\perp}\rangle$ versus $F_{D}$ for the the system in
Fig.~\ref{fig:1} with $F_{p} = 0.75$, 
$\alpha_{m}/\alpha_{d} = 1.25$, and $\theta^{\rm int}_{sk} = 51^\circ$.  
For $0 < F_{D} < 0.47$, the motion is strictly
along the direction of drive, with
$\langle V_{\perp}\rangle = 0$, and a series of  
sharp velocity drops appear in
$\langle V_{||}\rangle$.
This is region I, a shear flow phase, in which only the skyrmions in the
unpinned portion of the sample are moving.
For $0.45 \leq F_{D} < 0.72$,
these drops disappear and both
$\langle V_{||}\rangle$ and $\langle V_{\perp}\rangle$ increase with increasing $F_{D}$.
Here the sample has entered region III, a disordered plastic flow state,
where motion occurs both parallel and perpendicular to the drive, but some skyrmions
remain pinned and the overall skyrmion structure is disordered.
For $F_{D} > 0.72$, there is a change in the slope and the velocity in both directions
increases linearly with drive. 
This is region IV, where all the skyrmions
form a uniform lattice that moves in a direction that is close to the
intrinsic skyrmion Hall angle.

\begin{figure}
  \onefigure[width=\columnwidth]{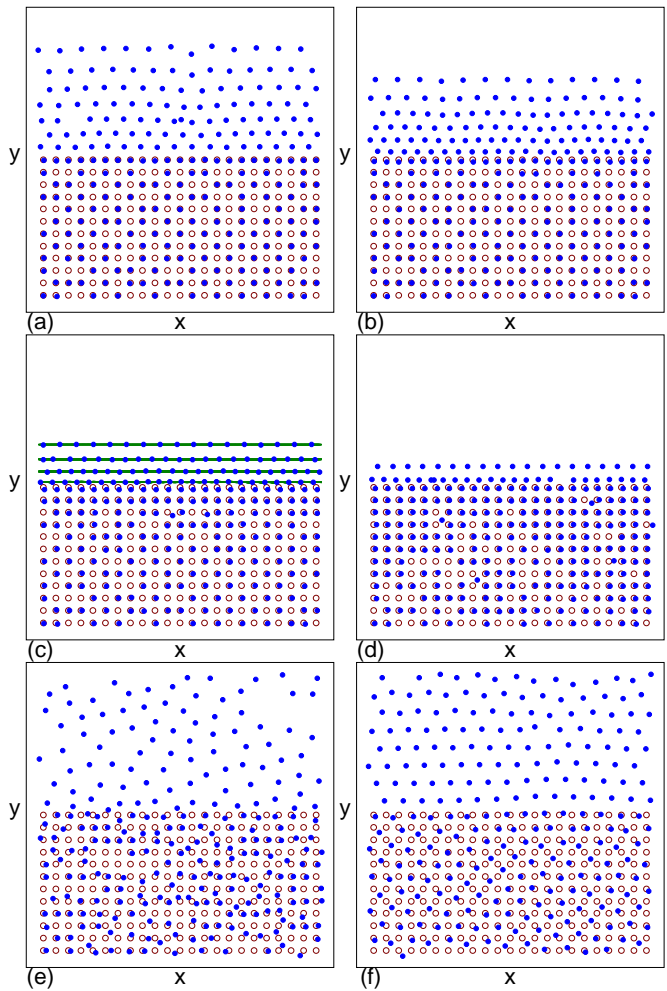}
\caption{
  Skyrmion positions (dots) and pinning sites (open circles) for the system in Fig.~\ref{fig:2}
  at the points marked a through f.
  (a) At $F_{D} = 0.125$,
  the top row of pinning sites next to the unpinned region is filled
  and a skyrmion density gradient starts to form in the unpinned region.
  (b) At $F_{D} = 0.225$, 
  there are six rows of moving skyrmions.
  (c) At $F_{D} = 0.325$, where the skyrmion trajectories (lines) are highlighted,
  there are four rows of moving skyrmions.
  (d) At $F_{D} = 0.425$, there are only two rows of moving skyrmions. 
  (e) $F_{D} = 0.6$ in the disordered plastic flow region III.
  (f) $F_{D} = 0.9$ in region IV or the moving lattice
state.
}
\label{fig:3}
\end{figure}

In Fig.~\ref{fig:1}(b) at $F_{D} =0.025$,  the skyrmions
in the unpinned region form a moving uniform triangular lattice.
As $F_{D}$ increases, the Magnus term pushes these skyrmions in 
the negative $y$ direction,
causing the top row of pinning sites to fill with skyrmions and producing
a density gradient of skyrmions
in the unpinned area, as shown in Fig.~\ref{fig:3}(a)
for $F_{D} = 0.125$.
Here
there are seven rows of
moving skyrmions in the unpinned region.
As $F_{D}$ is increased further, the skyrmions become   
more compressed and the number of moving rows  drops,
as shown in Fig.~\ref{fig:3}(b)
at $F_{D} = 0.225$  where there are six rows of moving skyrmions.
The sharp 
drops in $\langle V_{||}\rangle$ in Fig.~\ref{fig:2}
correspond to
drives at which the compression of the skyrmion structure generates
row reduction and partial 
row reduction events.
At $F_{D} = 0.275$ in Fig.~\ref{fig:3}(c), there are four rows of moving skyrmions,
as highlighted by the plot of the skyrmion trajectories.
After a series of additional small jumps in $\langle V_{||}\rangle$,
there are three rows of
moving skyrmions for $0.35 < F_{D} < 0.4$, and 
for $0.4 < F_{D} < 0.465$,
there are only two rows of moving skyrmions,
as shown in Fig.~\ref{fig:3}(d) at $F_{D} = 0.425$.
Once the drive is large enough,
the skyrmions in the pinned region begin to flow,
and the systems enters 
a disordered plastic flow state in which the skyrmions can move both
parallel and perpendicular to the drive.
Figure~\ref{fig:3}(e) illustrates
the skyrmion structure
in region III at $F_{D} = 0.6$.
At the highest drives, the system
enters a flowing lattice phase, as shown in Fig.~\ref{fig:3}(f) at $F_{D} = 0.9$. 
The moving lattice state is similar
to that observed at higher drives in simulations
with uniform pinning where the skyrmions
dynamically reorder into a crystal structure \cite{Reichhardt15a,Koshibae18}. 

The drops in the velocity-force curves indicate that
the system exhibits negative differential
conductivity with $d\langle V_{||}\rangle/dF_{D} < 0.$
Similar effects have been observed for vortices in periodic pinning arrays
at transitions from disordered two-dimensional flow to
effectively one-dimensional flow, where the number of moving vortices drops as the
drive increases \cite{Reichhardt97,Gutierrez09}.
In the vortex system, there is only one velocity drop,
while in the skyrmion system there are a series of drops.
The sudden changes in the skyrmion lattice
structure during compression are similar to what is found in
the compression of charged particles
by time dependent potentials with increasing confining strength,
where the compressed lattice
undergoes a combination of slow elastic deformations
interspersed with sudden drops due to plastic rearrangements,
and where the highest stability occurs when
an integer number of rows of particles can form a triangular lattice within
the confined region
\cite{McDermott16}.
In the skyrmion case, the compression is produced by
the increase in the Magnus force
with increasing drive that
pushes the moving skyrmions against the pinned region,
with
the pinned skyrmions acting as a barrier.
For varied filling ratios $N_s/N_p$,
we observe similar features in the velocity force curves,
and when the number of 
skyrmions is much
smaller than the number of pinning sites,
we find a drive-induced pinning 
phenomenon in which
all the skyrmions leave
the unpinned region
and become
trapped in the pinned region.

\begin{figure}
  \onefigure[width=\columnwidth]{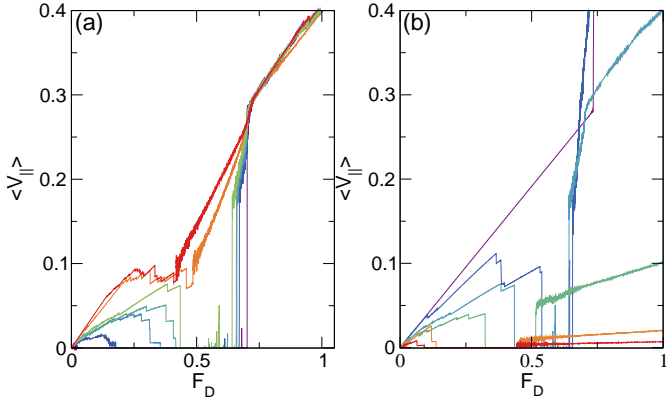}
\caption{
  (a) $\langle V_{||}\rangle$ vs $F_{D}$ for the system in
  Fig.~\ref{fig:2} at fixed $\alpha_{m}/\alpha_{d} = 1.25$ and $F_{p} = 0.75$ for
  $N_{s}/N_{p}= 1.1$ (red),
  $0.375$ (orange),
  $0.3125$ (light green),
  $0.25$ (blue green),
  $0.1875$ (light blue),
  $0.125$ (dark blue),
  and $0.0625$ (purple).
  Here there is a reentrant pinning region II
  for $N_{s}/N_{p} < 0.5$.
  (b) $\langle V_{||}\rangle$ vs $F_{D}$ for the same system
  at $N_{s}/N_{v} = 0.3125$ and
  $\alpha_{m}/\alpha_{d} = 0.0$ (purple),
  $0.75$ (dark blue),
  $1.25$ (light blue),
  $3.0$ (blue green),
  $9.0$ (orange),
  and $18.0$ (red).
  Here region II is lost for $\alpha_m/\alpha_d = 0$ but becomes
extended for increasing $\alpha_{m}/\alpha_{d}$.
}
\label{fig:4}
\end{figure}

In Fig.~\ref{fig:4}(a) we plot $\langle V_{||}\rangle$
versus $F_D$ for the same system as in Fig.~\ref{fig:2}
at varied skyrmion 
density of $N_{s}/N_{p}= 1.1$, $0.375$, $0.3125$, 
$0.25$, $0.1875$, $0.125$, and $0.0625$.
For $N_{s}/N_{p} = 1.1$ we find
the same phases I, III, and IV as in Fig.~\ref{fig:2},
but for $N_{s}/N_{p} < 0.5$ we observe a new reentrant phase, as shown
in the $N_{s}/N_{p} = 0.3125$
curve which has a transition from phase I flow to a state with
$\langle V_{||}\rangle = 0$ over the range $0.5125 < F_{D} < 0.575$.
In this same interval,
$\langle V_{\perp}\rangle = 0$ (not shown).
We label this reentrant pinning effect region II, and
as $N_{s}/N_{p}$ decreases, the onset of phase II shifts
to lower values of $F_{D}$. 

Since the reentrant phase arises
when the Magnus force pushes the skyrmions into the pinned region,
we also examine the effect of changing $\alpha_{m}$ while fixing
$\alpha_{d} = 1.0$. 
In Fig.~\ref{fig:4}(b) we plot $\langle V_{||}\rangle$ versus
$F_{D}$ for a system with $N_{s}/N_{p} = 0.3125$ 
at $\alpha_{m}/\alpha_{d} = 0.0$, $0.75$,  $1.25$,
$3.0$, $9.0$, and $18.0$.
In the overdamped case of
$\alpha_{m}/\alpha_{d} = 0.0$,
the drop in $\langle V_{||}\rangle$ disappears and the
velocity increases monotonically with
increasing $F_{D}$.
For $\alpha_{m}/\alpha_{d} < 0.35$  there is
no reentrant pinning phase since the skyrmion Hall angle is too small;
however, for $\alpha_{m} > 0.35$, phase II appears and becomes wider with
increasing $\alpha_m/\alpha_d$ 
as the skyrmion move more rapidly into the pinned region.
The onsets of phases III and IV also shift
to lower $F_{D}$ with increasing $\alpha_{m}$.

\begin{figure}
  \onefigure[width=\columnwidth]{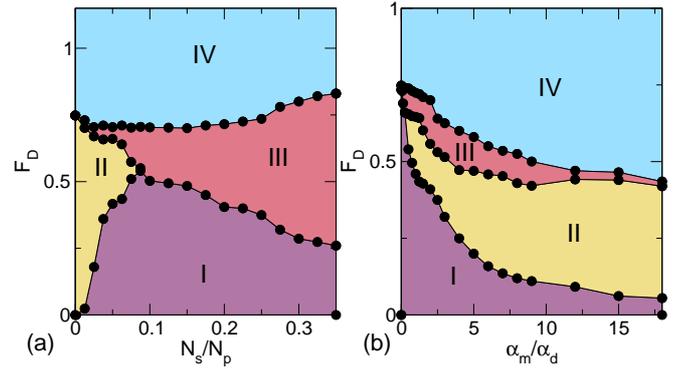}
  \caption{ (a) Dynamic phase diagram for the system in Fig.~\ref{fig:4}(a) as a
    function of
    $F_{D}$ vs $N_{s}/N_{p}$ at fixed 
    $\alpha_{m}/\alpha_{d} = 1.25$.
    I: shear flow (light purple); II: reentrant pinning (yellow);
    III: disordered plastic flow (pink); 
    IV: uniform moving crystal (blue).
    (b) Dynamic phase diagram for the same system
    as a function of $F_{D}$ vs $\alpha_{m}/\alpha_{d}$ at
    fixed $N_{s}/N_{p} = 0.3125$.
}
\label{fig:5}
\end{figure}

In Fig.~\ref{fig:5}(a) we plot the
dynamic phase diagram
as a function of $F_{D}$ versus the filling fraction $N_{s}/N_{p}$ for the system 
in Fig.~\ref{fig:4}(a) at $\alpha_{m}/\alpha_{d} = 1.25$.
Here, phase II appears only for
$N_{s}/N_{p} < 0.1$, while the width of
phase I decreases with increasing $N_s/N_p$.
The III-IV transition depends only weakly on the filling, and the
system can dynamically reorder when $F_{D}/F_{p} > 1.0$.
For the lowest filling of $N_{s}/N_{p} < 0.05$, 
region IV is a uniform moving liquid rather than
a moving lattice since the skyrmions are far enough 
apart that the skyrmion-skyrmion interactions are relatively weak. 
Figure~\ref{fig:5}(b) shows the
dynamic phase diagram as a function of $F_{D}$ versus $\alpha_{m}/\alpha_{d}$ in
the same system at $N_s/N_p=0.3125$,
highlighting the fact that the 
width of phase II increases
with increasing $\alpha_{m}$.

\begin{figure}
  \onefigure[width=\columnwidth]{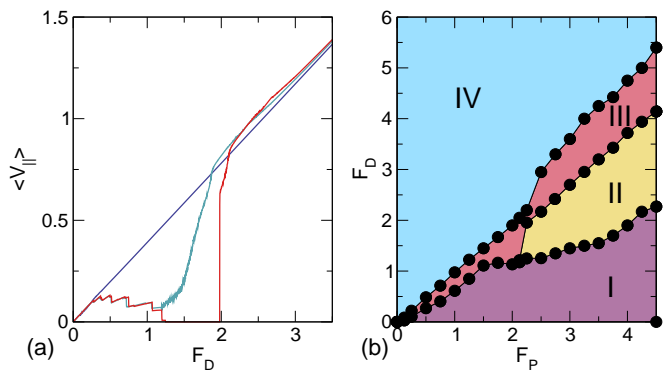}
  \caption{ (a) $\langle V_{||}\rangle$ vs $F_{D}$ for a system with
    $\alpha_{m}/\alpha_{d} = 1.25$ and $N_{s}/N_{p} = 1.0$ at
    $F_{p} = 0.25$ (dark blue), $2.0$ (blue green), and $2.25$ (red),
    showing that phase II can emerge for increased pinning strength. 
    (b) Dynamic phase diagram as a function of $F_{D}$ vs $F_{p}$
    for the system in (a) showing that phase II occurs when $F_{p} > 2.0$.
}
\label{fig:6}
\end{figure}

Phase II can occur for fillings $N_{s}/N_{p} > 0.5$
depending on the strength of the pinning. 
In Fig.~\ref{fig:6}(a) we plot $\langle V_{||}\rangle$ versus $F_{D}$
for a system with $\alpha_{m}/\alpha_{d} = 1.25$ 
at $N_{s}/N_{p} = 1.0$ with $F_{p} = 0.25$, $2.0$, and $2.25$,
showing that as $F_p$ increases,
the extent of region I grows,
and that when $F_{p} > 2.0$,
the reentrant pinning phase II can occur.
In Fig.~\ref{fig:6}(b) we
illustrate the dynamic phase digram as a function of $F_{D}$ versus $F_{p}$
for the system in Fig.~\ref{fig:6}(a), highlighting phases I-IV.
Reentrant pinning can occur when $F_{p} > 2.0$.
At $N_{s}/N_{p} > 1.0$, there is always some skyrmion motion in the
sample; however, there are still intervals of $F_{D}$
in which $\langle V_{||}\rangle$ is small. 
In general, for a system with either strong pinning or
large maximum forces on the skyrmions,
whenever the pinning is inhomogeneous
we would expect to find both the shear flow phase I and the
reentrant pinning phase,
along with an accumulation of skyrmions along the edges of
the pinned region.
Recent experiments have demonstrated drive-induced skyrmion
accumulation and skyrmion density gradients along the sample edges, and these
effects were attributed to
the Magnus force \cite{Sugimoto19}.
There has also been previous work
on drive-induced pinning effects for single skyrmions
interacting with magnetic pinning sites \cite{Muller15};
however, this is
a different effect than what we describe since in
Ref.~\cite{Muller15},
the pinning sites produce
a repulsive potential barrier that
the skyrmion must overcome
at finite drive in order to be captured.

In summary, we investigate skyrmion dynamics in a system with spatially
inhomogeneous pinning, where a square pinning array occupies only half of
the sample in a stripe aligned with the driving direction.
In the equilibrium state, the skyrmion density is uniform,
and under an applied drive,
skyrmion motion initially occurs only
in the unpinned region
and is in the same direction as the driving force.
As the drive increases, the skyrmions in the unpinned region
begin to form a density gradient
due to the Magnus force, with an accumulation
of skyrmions appearing at the edge of the pinned region.
A series of drops in the velocity-force curves occur when
sudden rearrangements of the moving skyrmions
occur as
the compression of the skyrmion lattice
increases with increasing drive.
The jumps are correlated with partial row
reduction events of the moving  skyrmions
as groups of skyrmions become trapped in the pinned region.
At higher drives, there is a transition to a disordered
plastic flow state followed by a transition 
to a moving lattice state,
in both of which the skyrmions move both
parallel and perpendicular
to the driving force due to the Magnus term.
When the number of pinning sites is sufficiently high,
a reentrant pinning phenomenon appears where
there is a finite drive at which all the skyrmions become trapped in the pinned
region.
The reentrant pinning regime increases in extent for increasing
intrinsic skyrmion Hall angle
since the skyrmions can move more rapidly
into
the pinning sites.
In the overdamped limit, the reentrant pinning and 
dynamical skyrmion lattice compression are lost. 
Our results indicate that inhomogeneous pinning can be used
to guide skyrmions,
and that the skyrmion gradient and accumulation
should be general features of skyrmion systems with inhomogeneous disorder.

\acknowledgments
This work was supported by the US Department of Energy through
the Los Alamos National Laboratory.  Los Alamos National Laboratory is
operated by Triad National Security, LLC, for the National Nuclear Security
Administration of the U. S. Department of Energy (Contract No. 892333218NCA000001).

\bibliographystyle{eplbib}
\bibliography{mybib}
\end{document}